\documentclass[12pt]{article}

\usepackage{graphicx}
\usepackage{amssymb}
\usepackage{psfrag}

\renewcommand{\=}{~=~}

\begin{document}

\begin{center}
   \textbf{\Large CINEMA, FERMI PROBLEMS, \& \\ GENERAL EDUCATION}\\[2mm]
       C.J. EFTHIMIOU\footnote{costas@physics.ucf.edu} and
       R.A. LLEWELLYN\footnote{ral@physics.ucf.edu}\\
  DEPARTMENT OF PHYSICS\\
  UNIVERSITY OF CENTRAL FLORIDA
\end{center}

During the past several years the authors have developed a new
approach to the teaching of \emph{Physical Science}, a general
education course typically found in the curricula of nearly every
college and university.  This approach, called \textit{Physics in
Films} \cite{EL}, uses scenes from popular movies to illustrate
physical principles and has excited student interest and improved
student performance. A similar approach at the high school level,
nicknamed \emph{Hollywood Physics}, has been developed by Chandler
Dennis \cite{Dennis1,Dennis2}. The two approaches may be
considered complementary as they target different student groups.

The analyses of many of the scenes in \textit{Physics in Films}
are a direct application of Fermi calculations --- estimates and
approximations designed to make solutions of complex and seemingly
intractable problems understandable to the student non-specialist.
The intent of this paper is to provide instructors with examples they can use to
develop skill in recognizing Fermi problems and making Fermi
calculations in their own courses.

\section{Fermi, Socrates, and Orders of Magnitude}

\subsection{In the Beginning there was Fermi…}

In the early morning of July 16, 1945, the first atomic bomb
exploded near Alamagordo, New Mexico.  Watching at the main
observation post a few miles from ground zero was Enrico Fermi
(1901--1954), the Italian physicist who had built the first
man-made atomic reactor a few years before.  About 40 seconds
after the bright detonation flash, as the air blast reached his
location he dropped a handful of torn bits of paper from a height
of about 1.5 meters above the ground.  There being no wind that
morning, he measured their displacement at the ground, about 2.5
meters, as the air blast passed him.  Doing a quick, approximate
calculation, he estimated the strength of that first atomic bomb
explosion to be equivalent to about 10,000 tons of TNT
\cite{Rhodes}. His quick calculation missed the actual strength of
the explosion, as measured by the instrumentation, by less than a
factor of two!

\begin{figure}[h!]
\begin{center}
\includegraphics[width=4cm]{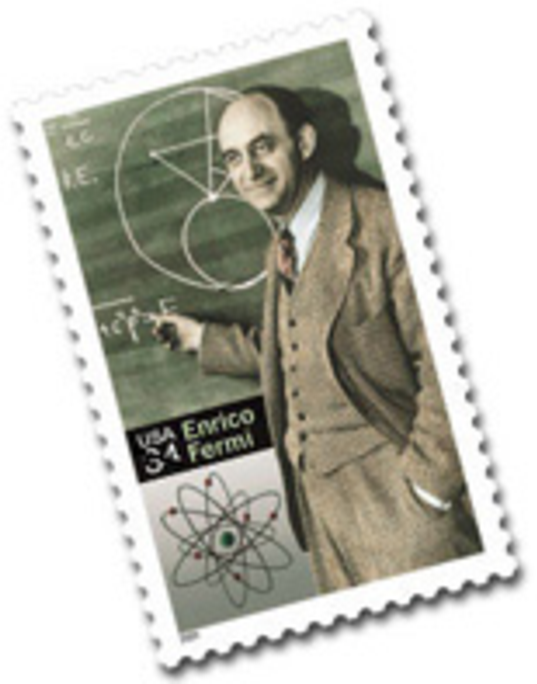}
\includegraphics[width=4cm]{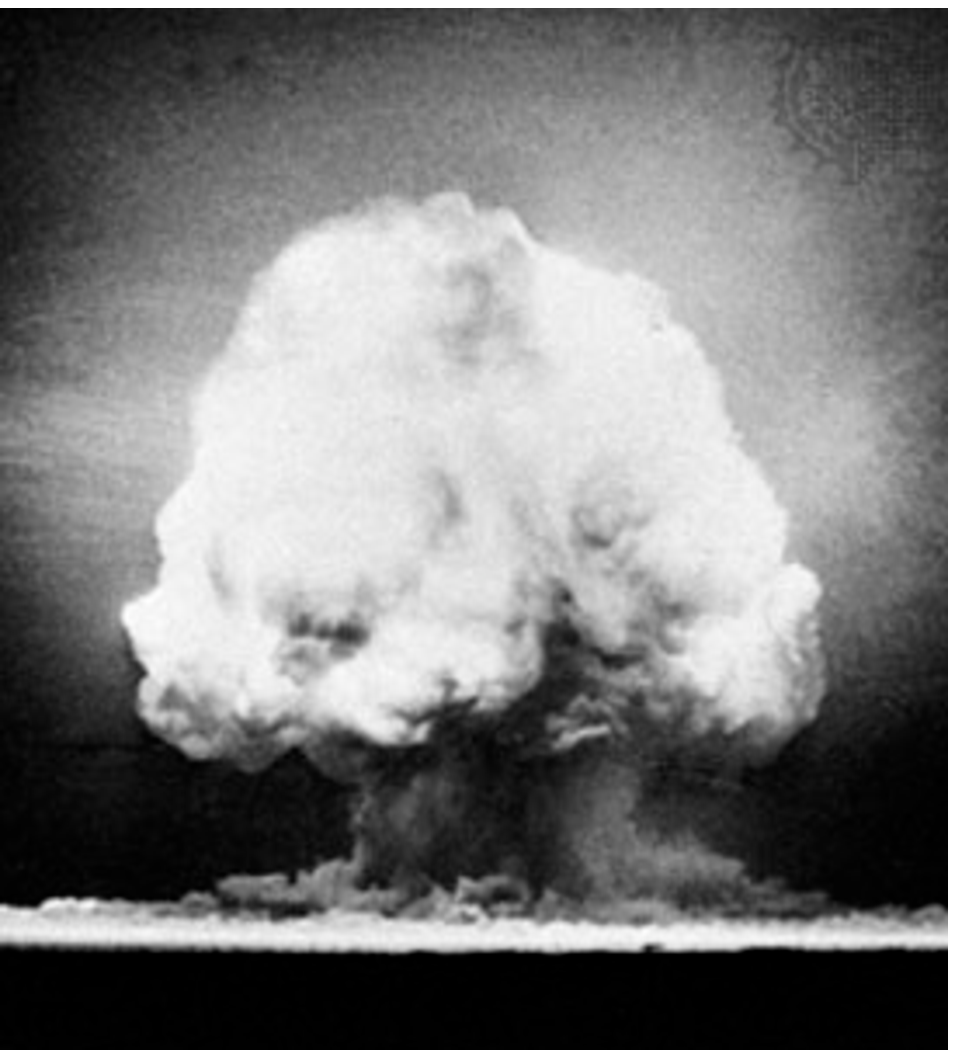}
\end{center}
\caption{\footnotesize Left: A US commemorative postage stamp
         which was issued at
         the University of Chicago on Sept. 29, 2001, portrays  Enrico
         Fermi. Right: First atomic bomb test, near Alamogordo, New Mexico,
         July 16, 1945. Picture is from U.S. Department of Energy, Los Alamos National
         Laboratory, New Mexico.}
\end{figure}

Fermi\footnote{A nice essay about Fermi problems can be found in
\cite{Baeyer}.} had an extraordinary ability to answer with
reasonable accuracy any question posed to him, questions that
would seem impossible to answer to an ordinary person. The classic
example of such questions that is attributed to him is `How many
piano tuners are there in the city of Chicago?' Asking this
question, even of  trained scientist, will initially create
frustration and a feeling that the answer may be unattainable, at
least without referring to the piano tuners' union website.
However, upon second, more careful thought, one discovers that the
question can be split into a series of simpler questions which
admit approximate answers leading eventually  to an approximate
(but very reasonable) answer to the original question:
  \begin{enumerate}
    \item What is the population of the Chicago?
    \item How many families does this correspond to?
    \item What fraction of families have pianos?
    \item What is average number of piano tunings per year per
          family?
    \item What is the average number of piano tunings per year
          that a tuner can make?
  \end{enumerate}
`Fermi problems' thus have a very distinct
profile: they  always seem vague with very little or no information given,
but they admit dissection into a set of simpler
questions that lead to the final answer. Once understood,
Fermi problems become a source of limitless fun. An answer found
is a blast of excitement and joy.

\subsection{Socrates' Dialectic Method}

Socrates (470--399 BC) was an Athenian philosopher and teacher of
Plato. One of the most significant philosophers of the western
world, he had (through Plato) direct influence on modern thought.
Socrates established the \emph{Socratic method} (\emph{dialectic
method}). The method is probably used by many---in not all---of
us.  Basically, through a series a questions and answers the
interviewer leads a person who thought he knew something to a
realization that he does not. Alternately, the series of questions
and answers is used to prepare the development of a more
sophisticated structure. Isn't this that we do all the time in the
classroom? Looking carefully at the concepts, Fermi Problems are a
variation of the Socratic method. The answer has to be extracted
from the person who was asked the Fermi Problem. However, in order
for him to answer the question, he has no other choice but asking
himself a series of questions that will help him reach a solution.
Indeed, the Socratic method is at the heart of the method of
solving Fermi Problems .

\subsection{Orders of Magnitude}

Physicists have a lifelong addiction to comparing
sizes\footnote{Size here refers to the magnitude of any physical
quantity; it is not restricted to the physical dimensions of an
object.}. Words such as `small' and `large' are useless in physics
in favor of more precise terms such as `small compared to' and
`large compared to'. The standard units---those of SI---are
usually scaled appropriately to make smaller and larger sizes. For
simplicity, prefixes that are multiples of 10 have been introduced
to standardize the system.  Table 1 lists the frequently used
prefixes.

\begin{table}[h!]
\begin{center}
\begin{tabular}{|c|c|c|} \hline
  PREFIX & SYMBOL   & VALUE       \\ \hline
  \dots  & \dots    & \dots        \\
  tera   & T        & $10^{12}$   \\
  giga   & G        & $10^9$      \\
  mega   & M        & $10^6$      \\
  kilo   & k        & $10^3$      \\
  hecto  & h        & $10^2$      \\
  deka   & da       & 10          \\ \hline
         &          & 1           \\ \hline
  deci   & d        & $10^{-1}$   \\
  centi  & c        & $10^{-2}$   \\
  milli  & m        & $10^{-3}$   \\
  micro  & $\mu$    & $10^{-6}$   \\
  nano   & n        & $10^{-9}$   \\
  pico   & p        & $10^{-12}$  \\
  \dots  & \dots    & \dots       \\ \hline
\end{tabular}
\end{center}
 \caption{\footnotesize Standard prefixes in SI.}
 \label{table:1}
\end{table}

By making use of the prefixes and standard units, physicists get a
feeling for the magnitude of a quantity which is often rounded off
to the nearest power of ten. For example the average life of a
human is of the order of $10^2$ years and the age of the universe
is of the order of $10^{10}$ years. A nice visualization of 39
orders of magnitude can be found in the website \cite{powers},
which is based on the classic film by Charles and Ray Eames
\cite{Eames}. A modern remake can be seen in the IMAX film
\emph{Cosmic Voyage} \cite{IMAX} that presents a narrated cosmic
zoom across 42 orders of magnitude.

 Real-life problems encountered by physicists are almost always hard and few
 admit analytical solution. Physicists, in an effort to understand the scales
 involved, approach them as Fermi Problems. Once a rough
 estimation is obtained and the range of the answer is known,
 educated decisions can be made as to which approaches and
 techniques are optimum for a more precise, but probably still
 approximate, solution.

An interesting example of a Fermi Problem that captures students'
attention in general education physical science and astronomy
courses is the question `How many technologically advanced
civilizations exist in our galaxy?' The question was first asked
and answered by astronomer Frank Drake; consequently, the equation
that estimates that number is known universally as \emph{Drake's
equation}. To answer Drake's question we need to approximate
\begin{enumerate}
    \item the number of stars in our galaxy;
    \item the fraction of stars that admit solar systems;
    \item the average number of planets in these solar systems
          that are in the habitable zone;
    \item the fraction of the planets that will develop life;
    \item the fraction of the planets in which life will involve
          to intelligent life;
    \item the fraction of the plants in which intelligent life
          will advance to create technological societies;
    \item the lifetime of such civilizations.
\end{enumerate}

\section{Why Fermi Problems in General Ed?}

According to surveys by the National Science Foundation
\cite{NSF}, while 70 percent of the U.S. population know that
Earth moves around the sun, only half know that it takes one year
to do so. Over half think that early humans lived at the same time
as the dinosaurs. Only about 42 percent know that electrons are
smaller than atoms and barely 35 percent know that the universe
started with a huge explosion.  Clearly, the traditional science
general education courses, particularly physical science, are not
doing their job of fostering science literacy.

In the summer of 2002 the authors began the \emph{Physics in
Films} version of the physical science course with goal of
improving the science literacy of the thousands of non-science
students who take the course at our institution each year.  In the
process of the continuing development of that program it has been
discovered that general education students, who normally shudder
at the thought of doing calculations of any kind, readily accept
and learn to emulate the Fermi calculation approach to dealing
with seemingly very difficult, if not impossible problems.  Here
is a quick and simple example that can help demonstrate the method
even at the first meeting of the class and at the same time use
content related to the class.

\subsection{Speed ...  of Earth!}

The concept of `speed' is one that arises early in physical
science.  Many students know that a world-class runner can sprint
$100 m$ in $10 s$ (or less), but have no clue about how fast,
i.e., at what speed the runner is moving.  Similarly, they know
that a hot sports car will accelerate to $60 mi/h$ ($97 km/h$) in
$3.5 s$, but don't how to find the average speed of the car.  We
find that by introducing them to the Fermi calculation approach to
determining a speed they initially see as impossible for them to
find, they gain confidence.  They discover that (a) they can solve
the simple problems like those above and (b) they retain more,
having `done it themselves'.

To find the speed of Earth, some numbers are needed, many of which
students generally may not know.  Table  \ref{table:2} contains
some useful data for Earth\footnote{George Goth has tabulated a
long list of  data \cite{Goth} that are useful in answering a very
wide range of Fermi problems.}. Earth's speed is given by:
\begin{equation}
   V_{Earth} \= { \mbox{distance~traveled} \over \mbox{time~required} }~.
\label{eq:1}
\end{equation}

In one year Earth completes one orbit around the sun, so the
distance traveled is the circumference $C$ of the orbit and the
time required is one year.
$$
 C \= 2\pi R \= 2\pi \times 150 \cdot 10^{6} km \= 9.42\cdot 10^8
 km~,
$$
where the radius R of Earth's orbit is given in Table
\ref{table:2}.

\begin{table}[h!]
\begin{center}
\begin{tabular}{|c|c|c|} \hline
 \textbf{Physical Quantity}           & \textbf{Magnitude}& \textbf{Units} \\ \hline
 Area of Earth's oceans (\% of total) & 70.8              & \%    \\ \hline
 Mass of Earth                        & $5.97\cdot10^{24}$& $kg$  \\ \hline
 Radius of Earth                      & $6.37\cdot10^6$   & $m$   \\ \hline
 Thickness of Earth's outer core      & $2.27\cdot10^6$   & $m$   \\ \hline
 Radius of Earth's inner core         & $1.2\cdot10^6$    & $m$   \\ \hline
 Radius of Earth's orbit around sun   & $1.5\cdot10^{11}$ & $m$   \\ \hline
\end{tabular}
\end{center}
 \caption{\footnotesize Some data for Earth for Fermi
         calculations.}
 \label{table:2}
\end{table}

Converting one year into hours provides another opportunity to
reinforce the treatment of units like numbers in multiplication
and division.
$$
   1 y = 1 y \times {365d\over 1y} \times {24h\over 1d} \= 8,760 h~.
$$
Now we can find Earth's speed from equation (\ref{eq:1}):
$$
   V_{Earth} \= { 9.42\cdot 10^8\over8,760}{km\over h}\= 108,000{km\over
   h}~.
$$
and the students are amazed!  During the hour they spent in class
they have traveled through space a distance of more than 2.5 times
Earth's circumference at the
equator!

Strictly speaking, the speed of Earth around the Sun, as presented
above, is a \emph{straightforward calculation}, not a Fermi
problem since we copied the Earth radius about the Sun---which is
the central quantity for the calculation---from an astronomical
table. However, the question becomes a Fermi problem if we require
that the Earth radius about the Sun cannot be read from any table
but, instead, should be derived from simple known facts. To this
end, one may, for example, use the well known fact that light from
Sun reaches Earth after 8 minutes. Therefore the distance between
Earth-Sun is 8-light minutes or about 144,000,000 kilometers. Of
course, using the travel time of light between Sun and Earth is
not the only piece of information that can be employed. Which data
are considered known in a Fermi Problem is arbitrary; the solver
is free to utilize all his knowledge and his ingenuity.


\section{Cinema Fermi Problems}

Since our goal is to train students in critical thinking and
reasoning, the Fermi Problems discussed in class are more
sophisticated than those in the previous section. Almost all of
them rely on information presented in a movie.  We call such
problems \emph{Cinema Fermi Problems}. The students have to
extract data from the spoken dialogue and the visual pictures.
Often the calculation is quite intricate, similar to those
encountered in a regular, algebra-based introductory physics
course. An example of such a calculation is given in \cite{EL}. In
that article the authors discussed the NASA plan in the movie
\emph{Armageddon} \cite{Armageddon} to use the explosion of a
nuclear bomb to split into two pieces an asteroid which is on a
collision course with Earth. The two halves will be deflected away
from Earth by the explosion. Based on the information given in the
movie and some additional reasonable estimates, it was shown that
the plan would only result in Earth colliding with---instead  one
asteroid---two smaller fragments, each about half the size of the
original asteroid hitting Earth just a few blocks apart. The
authors skip calculations that could also be performed (such as
three body calculations for the system moon-Earth-asteroid and the
tidal forces on Earth from the asteroid) or discussions of an
alternate real NASA scenario that has been proposed for such a
situation \cite{Lu}. Another Cinema Fermi Problem, based on the
movie \emph{Speed 2} has been described by the authors in
\cite{EL2}. There the authors explain that the deceleration of the
cruise ship that is crashing into the port is too small to have
the catastrophic effects on the passengers of the ship, contrary
to what is shown in the movie. Tretter \cite{Tretter} presents a
Cinema Fermi Problem that deals with scaling. Scaling arguments
are simple and powerful and known since Galileo's time
\cite{Galileo}. Below we present three additional \emph{Cinema
Fermi Problems} to help the reader become familiar with the
concept. For our course, we have screened hundreds of movies and
worked tens of such problems.

\subsection{Gravity on a space station}

Let's do a bit harder Fermi problem, `What is the artificial
gravity generated on a rotating space station?' One example that
the authors use in \emph{Physics in Films} is taken from the movie
\emph{2001: A Space Odyssey} \cite{2001}.  In the movie passengers
on board the space station live normally.  In scenes both inside
of and outside of the station, they walk about, prepare food, sit
in chairs, and there are no objects floating about as students are
accustomed to seeing in television coverage of NASA's orbiting
space shuttles and the International Space Station or the old Mir
space station.  To answer the question we need to know (a) the
rotational speed of the space station and (b) the radius of the
station. The film provides us scenes showing people (who we assume
are of average height), windows (whose size we estimate by
comparison with the people), and external views of the slowly
rotating space station (whose rotation period we can measure with
a wristwatch) that show the same windows that were seen from
inside. The size of the windows provides a way to estimate the
radius of the wheel-shaped space station.

\begin{figure}
  \begin{center}
  \includegraphics[width=6cm]{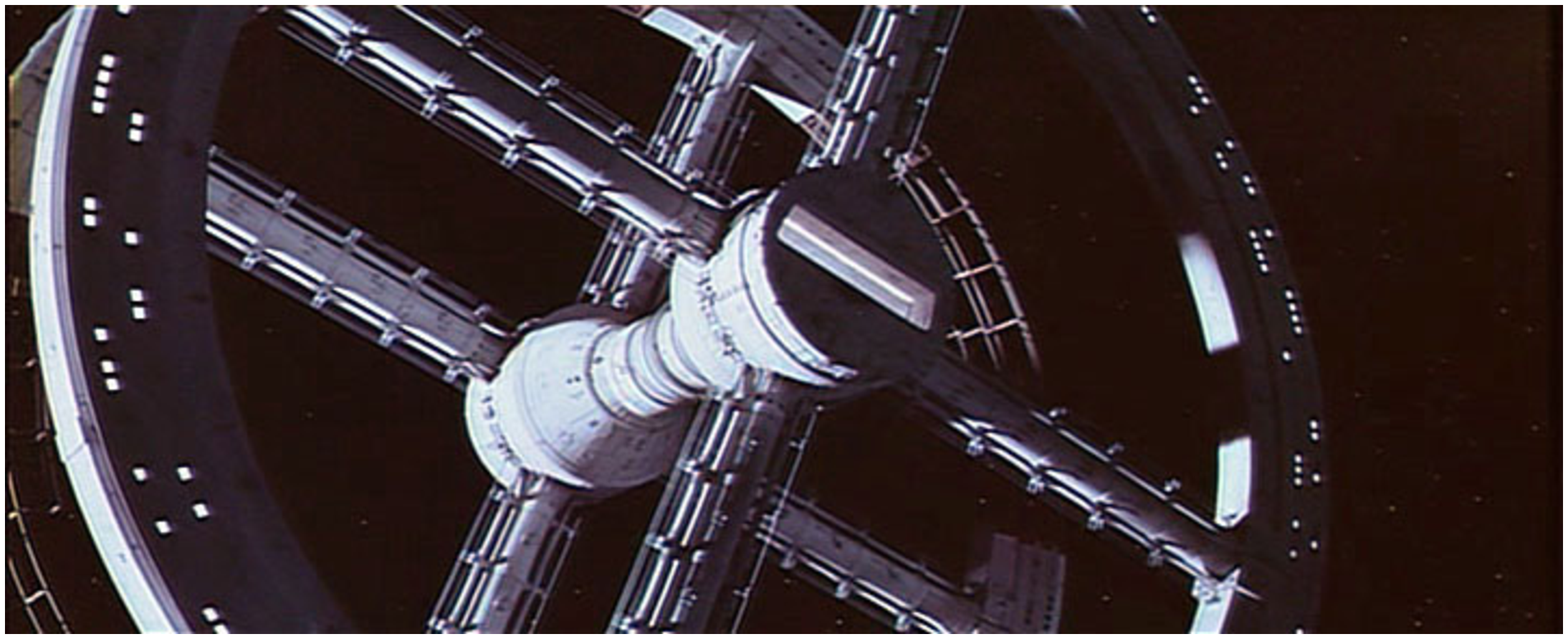}
  \hspace{1cm}
  \includegraphics[width=6cm]{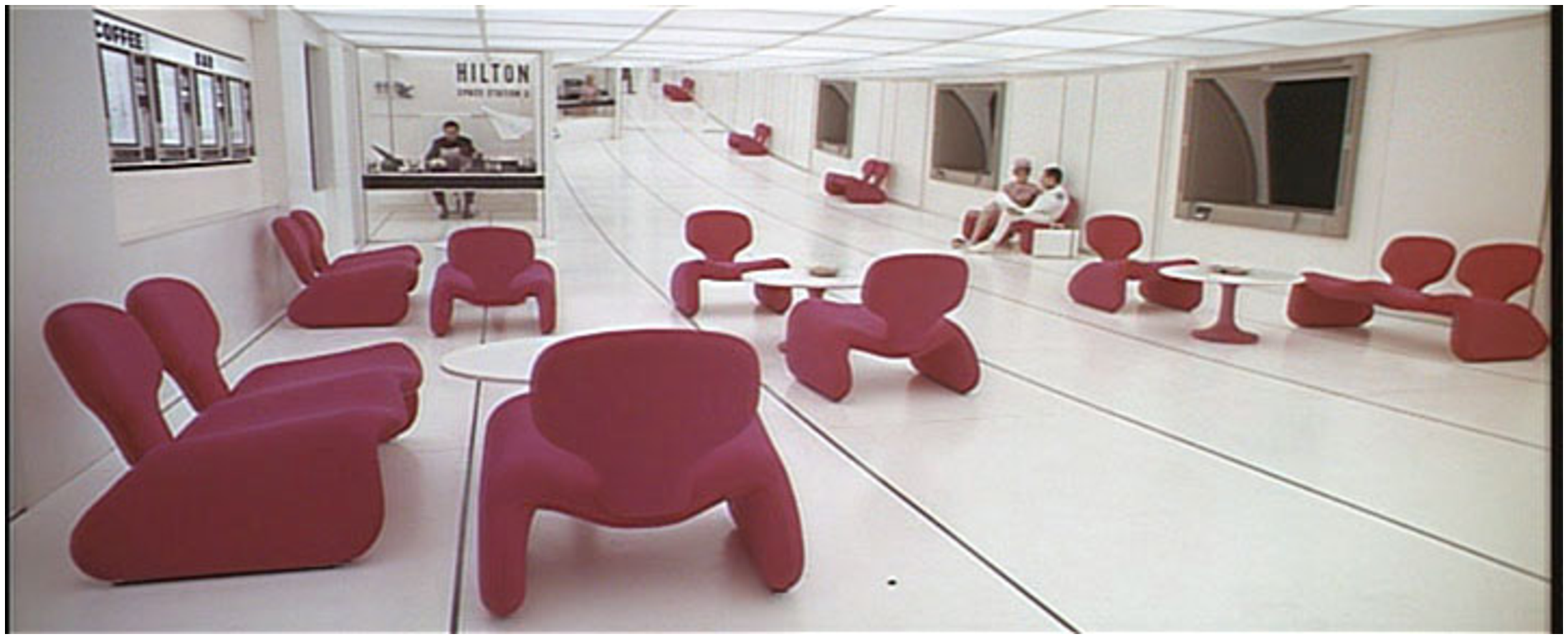}
  \end{center}
  \caption{\footnotesize Left: exterior view of the space station.
           Right: interior view of space station.}
  \label{fig:3}
\end{figure}

The film shows the rotation of the space station in real
time. This enables the computation of the time needed for one full
revolution.  Figure \ref{fig:3}  can be used in the estimation of
sizes. The left picture can be used to relate the radius of the
space station to the length of the windows. The second can be used
to compare the length of the window to an average person. From the
comparison of the lengths as explained above, we estimate that the
radius $r$ of the space station is about 300 meters.

From the movie, we see that one-fourth of a rotation occurs
in 9 seconds. Therefore, a full rotation of $2\pi$ radians occurs
in 36 seconds. This corresponds to an angular speed $\omega$ given
by
$$
   \omega \= {2\pi ~rad\over 36s} \= 0.174{rad\over s}~.
$$
Then, the centrifugal acceleration that is felt as gravity on the
outer 'rim' of the space station (where the people inside walk,
their heads toward the station's `hub') is equal to
$$
   a_{centrifugal} \= \omega^2r \= (0.174)^2\times300{m\over s^2}
                   \= 9.08 {m\over s^2}~,
$$
which is very close to $9.80 m/s^2$, the acceleration of gravity
on Earth.

Leaving this discussion, we present to our reader a related Cinema
Fermi Problem. The inside view of the space station (see Figure
\ref{fig:3}) clearly shows a curving floor. Is the curvature of
the floor consistent with the size of space station as shown in
the outside view?

\subsection{When a victory is worse than defeat}

\emph{Independence Day} \cite{ID4} became a Hollywood blockbuster
on the strength of its action and special effects. However, from
the physics point of view this is one movie that hardly makes
sense. In an effort to present the superiority of the aliens, the
director attributes to them spaceships of impressive dimensions
(both for the mothership and the battleships). These dimensions
are simply huge and the approach of such a ship to Earth  would
have serious effects that are not, of course, presented in the
movie. We shall not try to explore all problems raised by the
basic plot of the movie; instead, we will only demonstrate that,
under the premises of the director, humans will be obliterated.
That is, we shall present a simple calculation to show that, even
if humans successfully destroyed all of the battleships deployed
above the major cities on Earth, the ultimate result would be
holocaust for the human race.  Casting this as a Fermi Problem,
`What is the effect of destroying the alien space ship hovering
over a major city?  We need to know (a) how big is the ship; (b)
what is its density; (c) what is its mass; and (d) what is its
potential energy.

According to the movie, the battleships have a base diameter of 15
miles. This would imply a radius of 12 kilometers and a base area
$A=\pi R^2$ of about 452.4 square kilometers or $A=4.5\cdot10^{8} m^2$ in SI units.
To place this number in prospective, the borough of Manhattan in
New York City is about 59.5 square kilometers. Just one battleship would
cover about 8 Manhattans!

\begin{figure}[h!]
  \begin{center}
  \includegraphics[width=6cm]{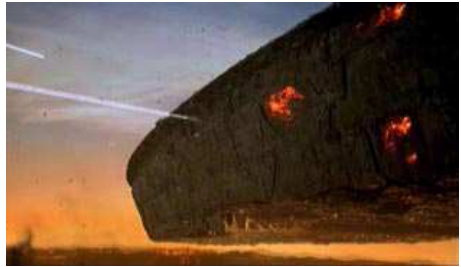}
  \end{center}
  \caption{\footnotesize A spaceship hovering above a city in \emph{Independence Day}.
           Ships of this size were deployed above every major city of the
           globe.}
  \label{fig:4}
\end{figure}

Watching the scene of the battle between the allied forces and the
ship, we can get a feeling of its height and also at the hovering
height above the city. Based on the stated diameter, we approximate the height $h$ of the
ship to be 1 kilometer and its hovering height $H$ about 2
kilometers.

\begin{table}[h!]
\begin{center}
\begin{tabular}{|c|c|c|} \hline
 \textbf{Physical Quantity}  & \textbf{Magnitude}  &  \textbf{Units} \\ \hline
 Density of water            &   $10^3$            &  $kg/m^3$ \\ \hline
 Density of Earth            &   $5.52\cdot10^3$   &  $kg/m^3$ \\ \hline
 Density of aluminum         &   $2.7\cdot10^3$    &  $kg/m^3$ \\ \hline
 Density of iron             &   $7.87\cdot10^3$   &  $kg/m^3$ \\ \hline
 Density of copper           &   $8.96\cdot10^3$   &  $kg/m^3$ \\ \hline
 Density of lead             &   $11.4\cdot10^3$   &  $kg/m^3$ \\ \hline
\end{tabular}
\end{center}
 \caption{\footnotesize Some density data for Fermi calculations.}
 \label{table:3}
\end{table}

From Figure  \ref{fig:4}, we see that the battleship has the shape
of a cylinder. Therefore, its volume would be $V=A\times h$ or
$V=4.5\cdot10^{11} m^3$. However not all of the volume of a
spaceship is material; a lot of it should be just empty space. In
order to estimate the mass of the ship, we must estimate how much
of the volume is material. A rough estimate would be about 10
percent. Therefore $V_{material}=4.5\cdot10^{10} m^3$. But to
estimate the mass of the ship, we also need to estimate the
density of the material. Based on our current experience,
spaceships are made of alloys. The density of metals is quite high
(See Table  \ref{table:3}). However, let us assume that we are
dealing with an extremely advanced species, far more advanced than
humans. It has mastered interstellar travel after all. So, it
seems reasonable, that the species has discovered a new material
of high strength but low density---close to the density of water,
$\rho=1000kg/m^3$. Therefore, the mass of the ship would be
$m=\rho\,V_{material}= 4.5\cdot10^{13} kg$. Hovering at a height
of 2 kilometers, the ship has stored $E=mgH=88.2\cdot10^{16}J$ of
potential energy. This energy will be released as heat after the
allied forces destroy the ship and it falls on the ground.
Recalling that the Hiroshima bomb released an amount of
$5\cdot10^{13}J$ of energy, the fall of the ship corresponds to
the detonation of 17,640 Hiroshima bombs! Remember, this happens
above every major city on Earth!

\subsection{Scientists who never studied Physical Science}

Probably, the worst movie that Hollywood has ever produced is
\emph{The Core} \cite{Core}. The director  worked hard to rewrite
the majority of the basic laws of physics. Eventually, he
succeeded. Like \emph{Armageddon} and \emph{Independence Day},
this movie can be the topic for numerous topics in physics. Here,
we shall concentrate only on the Army's plan to undo the problem
that it created: a secret project funded by the Army is
responsible for slowing the rotation of Earth's outer core. To
better understand this statement and its implications, a brief
explanation is in order.

\begin{figure}[h!]
  \begin{center}
  \includegraphics[width=6cm]{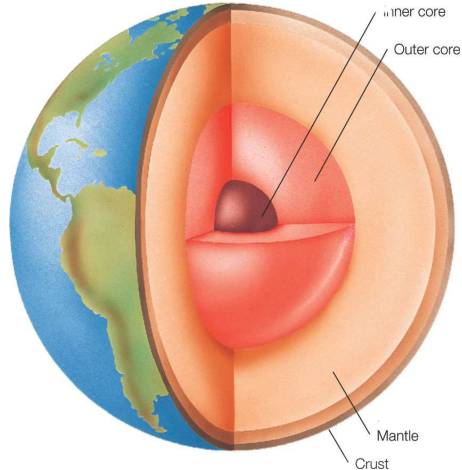}
  \end{center}
  \caption{\footnotesize The onion-like structure of Earth. Picture is
           borrowed from \cite{Tillery}.}
  \label{fig:5}
\end{figure}

Figure \ref{fig:5} shows a cross section of Earth. Earth has an
onion-like structure.  It is made of several shells: the crust,
the mantle, the outer core, and the inner core. We live on the
surface of the crust. But the other shells are as important as the
crust. The outer core is made of molten iron and rotates
relatively fast: it completes one full revolution in one day. This
gives the outer core an angular speed of
$$
  \omega_{outer} \= {2\pi~rad\over 1d} \= 7\cdot10^{-5}{rad\over
  s}~.
$$
The inner core is solid and rotates very slowly: only 1.5 degrees
a year. (This is equivalent to an angular speed of
$\omega_{inner}\sim 10^{-9}rad/s$.) Electric charges (ions and
electrons) circulating with the rotation of the iron in the outer
core are the likely source of Earth's  magnetic field. This field,
in turn, protects us from  harmful cosmic radiation (mostly
protons from the sun). In the movie, since the outer core has
slowed considerably and it is about to stop rotating, the effects
of cosmic rays on Earth start to be evident. Although we could
explore other obvious questions, such as, `How did the Army manage
to stop the core's rotation so effectively?'; `What happened to
the stored energy?', and `What else would happen to Earth if the
rotation of the outer core stops?', we shall only deal with the
plan to restore the rotation of the outer core. The scientists
have concluded that the only way to restore the rotation of the
outer core, is to carry a bomb of 1,000 megatons  to the outer
core and explode it. Leaving aside the question `How easy is it to
get there?', we will estimate how much rotation could be restored
with the bomb they used\footnote{The movie shows to us that the
bomb is placed \emph{inside} the outer core. As such, all forces
that develop are internal and one could argue that no rotation
could be created as a result of the explosion. However, the
director is lucky here. Internal forces, although they give a zero
net result, they can create a non-zero torque. For example, let's
look at the two-body system of the following figure.
  \psfrag{r1}{$\vec{r}_1$} \psfrag{r2}{$\vec{r}_2$}
  \psfrag{F}{$\vec{F}$} \psfrag{-}{$-$}
  \centerline{\includegraphics[width=4cm]{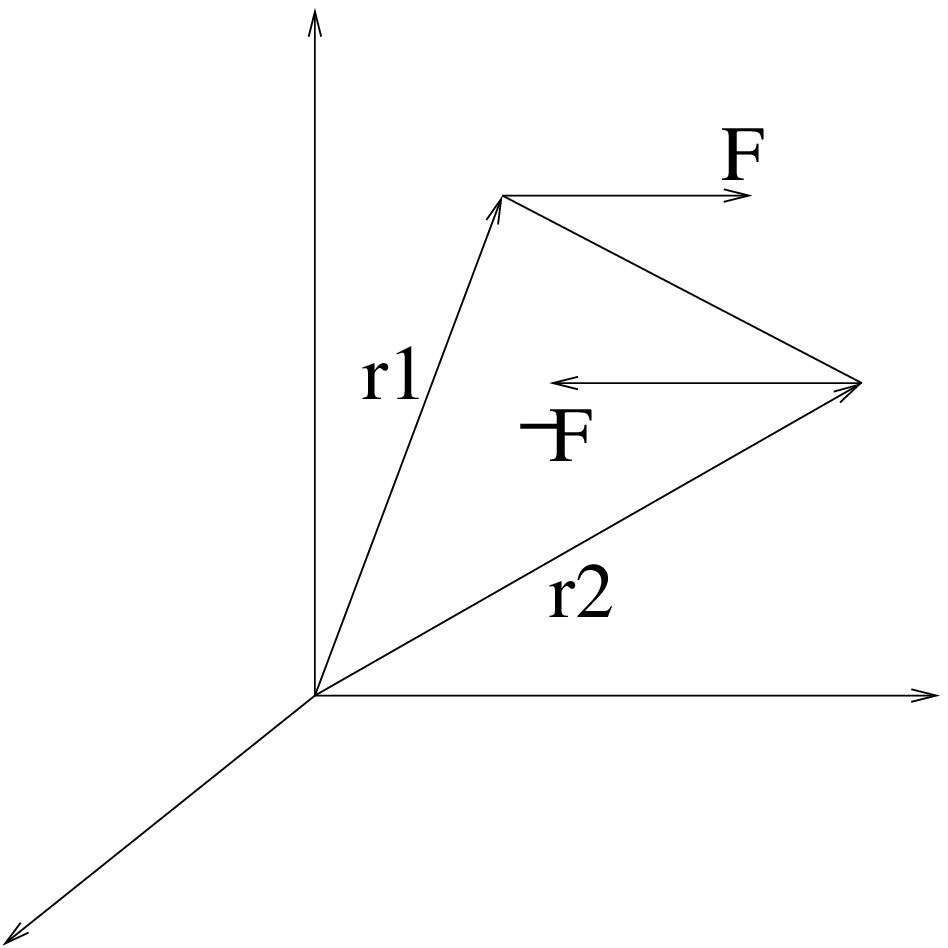}}
Obviously, the net torque is
$$
  \vec{\tau}_{net} = \vec{r}_1\times \vec{F} -\vec{r}_2\times\vec{F}
             = (\vec{r}_1-\vec{r}_2)\times \vec{F}~.
$$
The net torque will vanish only if the interaction between the two
particles of the system is along the line that joins the
particles. This is true for most cases but not necessarily all
situations.  We will thus give the director the benefit of doubt
and assume that releasing the bomb inside the outer core is
fine.}.

The Hiroshima bomb was 12 kilotons and released $5\cdot10^{13}J$
of energy. Therefore, the bomb carried to the outer core will
release approximately $E=4.2\cdot10^{18}J$. Assuming that
all this amount will become rotational energy of the outer core
(of course, not true), the angular speed given by the the
explosion is found by the equation:
$$
  E \= {1\over2}\, I_{outer} \, \omega^2~,
$$
where $I_{outer}$ is the moment of inertia of the outer core. This
can be computed very easily since inertia is an additive quantity
and the outer core is a spherical shell: If $R_o, R_i$ are the
radii of the surfaces of the outer and inner cores respectively and
$I_o, I_i, M_o, M_i$ stand for the moments of inertia and masses
for the spheres with radii $R_o, R_i$ respectively, then
 \begin{eqnarray*}
   I_{outer} &=& I_o -I _i
             \=  {2\over5} M_o R_o^2 - {2\over5} M_i R_i^2~.
 \end{eqnarray*}
If $\rho$ stands for the density of the outer core, then
$M_o=\rho\, ({4\pi/3}) \, R_o^3$, $M_i=\rho\, ({4\pi/3}) \,
R_i^3$, and
 \begin{eqnarray*}
   I_{outer} \=  {8\pi\over15}\,\rho\, (R_o^5 - R_i^5)~.
 \end{eqnarray*}
 It is known that $R_o=3470km$ and $R_i=1220km$ and $\rho=7870kg/m^3$
 (see tables \ref{table:2} and \ref{table:3}).
 Substituting all numbers in the equations we find
 $$
     \omega ~\simeq~ 10^{-9}{rad\over s}~.
 $$
 But this is 70,000 times smaller than the original angular speed!
 So, using a bomb that is equivalent to 83,000  Hiroshima
 bombs is not going to make any difference. They should have
 carried a bomb that was equivalent to 406,700,000,000,000
 Hiroshima bombs. Recall that this assumes that all energy
 given off by the explosion will become rotational energy. And, ... as
 always, the scientists who did the calculations were the best
 humanity has ever produced!  A Fermi Problem calculation revealed the absurdity of this plan.

\subsection{What seems obvious is not always right}
We are familiar with human motion on Earth. We are also familiar
with human motion on the Moon. From the many pictures NASA has
popularized, almost everyone knows that the Moon has low gravity
and walking is hard there. Instead, leaping is quite effective.

\begin{figure}[htb!]
  \begin{center}
  \includegraphics[width=6cm]{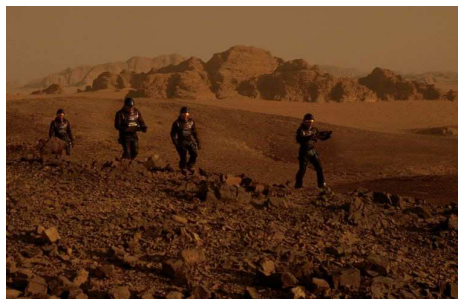}
  \end{center}
  \caption{\footnotesize A team of people walks on the surface of Mars as it would walk
           on Earth. Still is taken from \emph{Red Planet} \cite{RedPlanet}.}
  \label{fig:onMars}
\end{figure}

Hollywood has also popularized a planet Mars where everything
happens as usual. Recently, astronomers have been talking about Mars and
drawing attention to the fact that Mars is similar to Earth. In
fact, in the distant past Mars was very much like Earth with oceans
of water and the possibility of harboring life. Hollywood has
taken the opportunity and has produced a number of movies where
life on Mars looks identical to Earth. (By this, we really mean
the mechanics and response of the human body to Martian gravity.)
As a Cinema Fermi Problem, we
would like to determine how close to reality Hollywood's
depiction of Martian living has been.

\begin{table}[h!]
\begin{center}
\begin{tabular}{|c|c|c|} \hline
 \textbf{Physical Quantity}  & \textbf{Magnitude}    &  \textbf{Units} \\ \hline
 Mass of Moon                & $7.35\cdot10^{22}$    & $kg$ \\ \hline
 Radius of Moon              & 1,740                 & $km$ \\ \hline
 Mass of Mars                & $6.42\cdot10^{23}$    & $kg$ \\ \hline
 Radius of Mars              & 3,400                 & $km$ \\ \hline
\end{tabular}
\end{center}
 \caption{\footnotesize Some data for Fermi calculations.}
 \label{table:4}
\end{table}

The acceleration of gravity on a planet (or satellite) is given by
the simple formula
$$
    g \= G\, {M\over R^2}~,
$$
where $G$ is Newton's universal constant of gravity, $M$ and $R$
are the mass and radius of the planet (or satellite), respectively.
We can apply the same formula for Earth
$$
    g_{Earth} \= G\, {M_{Earth}\over R_{Earth}^2}~,
$$
where $g_{Earth}$ is the well known $9.80m/s^2$. Dividing the
previous two formul\ae, and solving for $g$, we find:
$$
   g \= g_{Earth} \, {M \over M_{Earth}} \, \left({R_{Earth} \over
   R}\right)^2~.
$$
Using tables \ref{table:2} and \ref{table:4}, we find
\begin{eqnarray*}
    g_{Moon} &=& 0.17 \, g_{Earth} \= 1.67 {m\over s^2}~, \\
    g_{Mars} &=& 0.38 \, g_{Earth} \= 3.72 {m\over s^2}~.
\end{eqnarray*}
The acceleration of gravity on Mars is about twice that on the
Moon, but still only 40 percent of that on Earth. Locomotion on Mars cannot
be identical to that on Earth. Here is a quick way to understand
this. Let $V$ be the speed of walking of a human in a
gravitational field $g$ and $L$ is the length of his leg. Treating
the leg as a solid rod that rotates about its one end, the (centripetal)
acceleration at the other end is $V^2/L$. However, this
acceleration can be at most $g$ (since the motion happens under
the influence of gravity only). Then
$$
   {V^2\over L} \le g ~\Rightarrow~ V \le \sqrt{gL}~.
$$
The length of the the leg of a human adult is of the order of
$1m$. Therefore, the maximum walking speed is
\begin{eqnarray*}
     V_{max} &=&  3.13{m\over s}~,~~~\mbox{on~Earh}~, \\
     V_{max} &=&  1.93{m\over s}~,~~~\mbox{on~Mars}~, \\
     V_{max} &=&  0.62{m\over s}~,~~~\mbox{on~Moon}~.
\end{eqnarray*}

\section{Concluding Comments}

There is a growing concern among scientists of the effect the
entertainment industry has on the public \cite{Sparks}. Since not
everyone who watches movies and TV series perceives the presented
material as fictional and an artistic creation for entertainment
only, the unchallenged presentation of pseudoscientific and
scientifically inaccurate topics by this industry creates and reinforces
unfound beliefs that threaten the
scientific literacy of the society.

Even worse than the spread of pseudo and incorrect science is the
negative stereotype of scientists that the entertainment industry
has created \cite{Evans,Tudor,Frayling} and how willing it is to
defame them in the name of quick profit. In the words of Evans
\cite{Evans}:
\begin{quote}
        \hspace{4mm} Popular entertainment media have long portrayed
        scientists as mad, bad, and dangerous to know, but in the
        past few decades entertainment media portrayals of science
        have changed significantly, and these changes seem to have
        accelerated in recent years. Science remains dangerous,
        but it is also increasingly portrayed as useless in
        solving problems. The skepticism about paranormal claims
        that is a part of scientific thinking is portrayed as a
        handicap. And in many newer entertainment media offerings
        the paranormal is portrayed as, well, normal ...

        \hspace{4mm} Film and television entertainment programming
        increasingly portrays science and reason as tools that are
        unsuitable for understanding our world in a new age of
        credulity ...

        \hspace{4mm}... Our entertainment mass  media provide a
        steady diet of negative images of science and
        skepticism---images that reflect and reinforce popular
        misgivings and misunderstandings about science.
\end{quote}

Given such a harsh antiscientific environment, it seems natural
for the public to develop a strong negative impression of science and many
young people to consider it an undesirable career. Our course
\emph{Physics in Films} is an effort to start challenging the
material presented to the public by the entertainment industry and
eventually reverse the observed trend of declining scientific
literacy. The use of Fermi problems and calculations in the way
demonstrated in this article has proven to be a significant factor
in students' improved understanding of the principles of physical
science \cite{EL,EL3}. Even though at the beginning they could not
individually perform the analyses nor, for most of them, the
algebra involved, they quickly learn to follow the mathematical
arguments and begin to think critically regarding other scenes in
the movies.  Our hope and expectation is that they will extend
this new-found ability beyond the classroom.


\end{document}